\RequirePackage{fix-cm}
\documentclass{svjour3}                     % onecolumn (standard format)
\smartqed  % flush right qed marks, e.g. at end of proof
\usepackage{graphicx}
\usepackage{graphicx}
\usepackage{amsmath}
\usepackage{amsfonts}
\usepackage{dirtytalk}
\usepackage{csvsimple}
\usepackage{enumitem}
\usepackage{cite}
\usepackage{adjustbox}
\usepackage{float}
\usepackage{apacite}
\usepackage{algorithm2e}
\usepackage{url}
\usepackage{xcolor}
\usepackage{setspace}
\usepackage{rotating}
\usepackage{caption}
\usepackage{textcomp}
\usepackage{cite}
\usepackage{amssymb}
\usepackage{multicol}
\usepackage{tabularx}
\usepackage[flushleft]{threeparttable}
\usepackage[margin=1.3in]{geometry}
\usepackage{apacite}
\usepackage[labelsep=endash]{caption}
\usepackage{ragged2e}
\usepackage{extdash}
\RestyleAlgo{ruled}

\begin{document}

% \title{\centering{\Large{\textit{Design and implementation of a standardized framework to allow the use of both observational and randomized data for precision medicine purposes}}}\setstretch{1.5}}

\title{The  R.O.A.D. to precision medicine}

\author{Dimitris Bertsimas$^{1}$, Angelos Koulouras$^{1}$, Georgios Antonios Margonis$^{*1,2}$
}

% \authorrunning{Short form of author list} % if too long for running head

\institute{Dimitris Bertsimas \at
              \email{dbertsim@mit.edu} 
           \and
           Angelos Koulouras \at
              \email{angkoul@mit.edu} 
            \and
           Georgios Antonios Margonis, MD, PhD; (Corresponding Author) \at
              \email{margonig@mskcc.org}
            \and
            1  Sloan School of Management and Operations Research Center, E62-560, Massachusetts Institute of Technology,
                MA, 02139 \\
                \\
            2  Department of Surgery, Memorial Sloan Kettering Cancer Center, New York, NY, 10065
}

\date{Received: NA / Accepted: NA}

\maketitle

\begin{abstract}
We  propose a prognostic stratum matching framework that addresses the deficiencies of \textbf{R}andomized trial data subgroup analysis and transforms \textbf{O}bserv\textbf{A}tional \textbf{D}ata to be used as if they were randomized, thus paving the road for precision medicine. Our approach counters the effects of unobserved confounding in observational data by correcting the estimated probabilities of the outcome under a treatment through a novel two-step process. These probabilities are then used to train Optimal Policy Trees (OPTs), which are decision trees that optimally assign treatments to subgroups of patients based on their characteristics. This facilitates the creation of clinically intuitive treatment recommendations. We applied our framework to observational data of patients with gastrointestinal stromal tumors (GIST) and validated the OPTs in an external cohort using the sensitivity and specificity metrics. We show that these recommendations outperformed those of experts in GIST. We further applied the same framework to randomized clinical trial (RCT) data of patients with extremity sarcomas. Remarkably, despite the initial trial results suggesting that all patients should receive treatment, our framework, after addressing imbalances in patient distribution due to the trial's small sample size, identified through the OPTs a subset of patients with unique characteristics who may not require treatment. Again, we successfully validated our recommendations in an external cohort.  
\end{abstract}

\section{Introduction}

Before the advent of evidence-based medicine (EBM), clinicians primarily based their decision making on prior experience as well as pattern recognition of previously encountered pathology. Thus, the longer a clinician had been in practice, the more diverse the pool of observed patient characteristics from which they could draw; this may explain why professional experience has historically been more valued in medicine than in any other field. This approach is somewhat intuitive in that biological heterogeneity of patients will inevitably present in unique ways that require individualized treatments. However, this approach is also extremely subjective (clinicians often cannot explicitly explain their own decision making), prone to cognitive biases, and cannot be assessed for validity and generalizability \cite{weed1999opening, kahneman2011thinking}.

The advent of EBM addressed these limitations but also created a paradox. Specifically, recommendations that stem from EBM, such as findings from randomized clinical trials (RCTs), may be suitable for the average patient but not for those who diverge from the norm \cite{deaton2018understanding}. Of note, Sir Austin Bradford Hill, who is the “father” of randomized clinical trials, admitted that trials do “not tell the doctor what he wants to know” and acknowledged that we need to “identify the individual patient for whom one or the other of the treatments is the right answer” \cite{hill1966reflections}.

Thus, there is a need to identify a way to understand how treatment effects can vary across patients, a concept described as heterogeneity of treatment effects (HTE). The practical application of this knowledge would allow for precision medicine, which is the optimal matching of the available treatments to each patient subgroup (or reference class). However, despite numerous publications on this topic, no methodology has gained considerable traction among physicians.

We believe there are four primary challenges. The first is the presence of confounding bias, especially with observational data. To counter this bias, one may argue that only randomized trial data should be used, but observational data are abundant and inexpensive whereas trials are commonly not feasible due to reasons such as disease rarity, high cost, or even ethical issues. Confounding in observational data has been widely acknowledged, and some matching methodologies have managed to at least partially balance the observed confounders \cite{stuart2010matching}. However, these approaches cannot balance unobserved confounders and thus function under the assumption that unobserved confounders do not exist (i.e., the “unconfoundness” assumption) \cite{hernan2021methods, athey2016recursive, wager2018estimation}. Of note, this assumption is dangerous to make in the medical field given our incomplete knowledge of the numerous biological factors that determine patient outcomes.

Second, previous approaches typically lacked an output in the form of a clear recommendation for clinicians. For example, a multidisciplinary technical expert panel issued the “Predictive Approaches to Treatment effect Heterogeneity” (PATH) statement on how to use regression-based prediction to facilitate precision medicine \cite{kent2020predictive}. The authors used data from randomized controlled trials to avoid unobserved confounding and presented a methodology that estimates the relative and absolute treatment effects across the various risk strata. The problem with this approach is that the risk strata are probabilities and not patient characteristics. Thus, clinicians can neither assess this approach for clinical intuitiveness nor can they use it to define patient subsets, which is a core pillar of precision medicine. Other approaches that have been used in the medical field to identify HTEs are causal forest machine learning (ML) algorithms \cite{venkatasubramaniam2023comparison}. Although they can detect heterogeneity in individual treatment effects, they can only report the relative feature importance of the predictors of differential treatment effects and do not output patient subsets defined by specific characteristics \cite{venkatasubramaniam2023comparison}. Notably, one relatively new approach, the SHAP visualization, can output the HTE of patient subsets defined by specific characteristics \cite{lundberg2017unified}. The disadvantage of this method is that it can only screen for interactions between one characteristic and the treatment. Thus, it cannot be used to find patient subgroups that are defined by more than one characteristic.  This is problematic because individual patients may differ from one another across many variables simultaneously, and the combination of these variables defines subgroups with different HTEs.

Third, the validation of the output of these approaches is problematic. Specifically, for the approaches that do not define patient subgroups based on their characteristics, validation can only concern either the model per se or the arbitrary strata (e.g., quintiles). For example, the validation of the output of an extension of the aforementioned PATH in observational data was done by calculating and comparing the magnitude of HTE in each of the strata across different datasets \cite{rekkas2023standardized}. In comparison, for the few approaches that define patient subgroups based on their characteristics (e.g., SHAP visualization), a cox proportional hazard model is typically trained in an external or internal test cohort to assess the statistical significance of the association of a treatment with prognosis in each of the patient subgroups \cite{zarinshenas2022machine}. The problem is that a statistically significant association of a treatment with prognosis does not inform clinicians about the percentage of patients who would receive an unnecessary treatment if this recommendation to treat was followed. Conversely, a non-significant association of a treatment with prognosis does not inform clinicians about the percentage of patients who would not receive a beneficial treatment if this recommendation to not treat was followed. 

Lastly, although randomized trial data do not share the constraints of observational data, the conventional subgroup analyses that are typically used to identify HTE in randomized trial cohorts may lead to grossly misleading results. Specifically, the one-variable-at-a-time subgroup analyses of RCTs involve sequentially categorizing patients based on single characteristics recorded at baseline (e.g., male vs female; old vs young) and testing whether the treatment effect varies across these categories.  This approach often neglects the fact that patients may differ from one another across multiple variables simultaneously. Additionally, power-related issues further diminish the reliability of these subgroup analyses, typically producing results that lack credibility, with many apparent positive subgroup effects ultimately proving to be inaccurate or exaggerated \cite{brookes2001subgroup, brookes2004subgroup}.

The primary objective of this paper is to propose and demonstrate, using two clinical examples, a framework designed to rectify the significant shortcomings in both a) the HTE analysis of observational data and b) subgroup analysis of randomized trial data.
Specifically, we employ a prognostic stratum matching framework and counter the effects of unobserved confounding in observational data by correcting the estimated probabilities of the outcome under a treatment through a novel two-step process. These probabilities are then used to train Optimal Policy Trees which are decision trees that optimally assign treatments to subgroups of patients based on their characteristics. This approach enables the creation of clinically intuitive treatment recommendations. We applied our framework to observational data of patients with gastrointestinal stromal tumors (GIST).  Using sensitivity and specificity as the relevant metrics we validated the OPT recommendations and showed that they outperformed those of experts. 
We also applied this framework to an RCT cohort of patients with extremity sarcomas, where the small sample size caused imbalances in the distribution of treated and untreated patients in the high-risk stratum. Our framework can identify and rectify such imbalances, which allowed the OPT to identify a subset of patients with distinct characteristics who may not require treatment. This has significant clinical implications as it challenges the current recommendations for clinical practice. The OPT recommendations were successfully validated in an external cohort.

More broadly, we believe  the proposed framework has the potential to revolutionize medical research. Furthermore, as an open access methodology it may bring a) effectiveness to research by equipping scientists worldwide with a "tool" that demands only a fraction of the time, expenses, and data compared to RCTs b) fairness by including women and minorities, who are historically underrepresented in RCTs, c) full use of RCT data instead of only using them to derive the average treatment effect (ATE), and d) the ability to generate knowledge in areas where RCTs are not possible.

\section{Methods}\label{sec1}

In this section, we establish the theoretical foundation of our framework by using an example. We focus on an observational dataset involving patients with gastrointestinal stromal tumors (GIST) who are undergoing post-surgery follow-up and can receive either imatinib treatment or observation. Our objective is to allocate treatments to patients in a manner that minimizes the risk of recurrence.

Suppose that we have $n$ patients with GIST in our observational dataset. Each patient $i$ is characterized by a vector of covariates $x_{i}$, such as mitotic count, tumor site, and tumor size. Let $t$ denote the treatment in general and $t_{i}$ denote the treatment of patient $i$. We set $t_{i}=1$ if the patient received imatinib; otherwise $t_{i}=0$. The outcome $y_{i}=1$ if the patient had a recurrence; otherwise $y_{i}=0$. Thus, the characteristics of patient $i$ are summarized by $(x_{i}, t_{i}, y_{i})$ for $i=1, \cdots, n$. 

As this is an observational dataset, treatment was not randomly assigned and patients who were predicted to have a worse baseline risk of recurrence were more likely to be offered imatinib by their physicians. Thus, since $t_{i}=1$ patients have a higher baseline risk of recurrence (i.e., the risk of recurrence if no treatment is offered), a direct comparison of outcomes in the $t=1$ and the $t=0$ groups is a suboptimal way to estimate the effects of imatinib. Indeed, in this example, the mean outcomes of the patients who received imatinib were worse than the mean outcomes of patients who did not. Since imatinib is not detrimental (i.e., does not increase the risk of recurrence), this indicates that confounding bias is present in this dataset.\cite{laurent2019adjuvant} 

The source of confounding bias is factors that increase both the baseline risk of an outcome and the chances that a treatment to prevent the outcome is given. In this example, greater mitotic count, a non-gastric tumor site, and greater tumor size are observed confounders as they are associated with both an increased risk of recurrence and a higher likelihood of receiving imatinib. Although it is possible to match for these observed confounders between the $t=1$ and $t=0$ groups to balance their effect in the two groups, other confounders likely exist. Importantly, such confounders may not be observed in the dataset as they have not yet been discovered, which means that we cannot balance their effects by matching. To mitigate this issue, we propose a two-step approach.

\subsection{Risk Strata and Matching}

First, we train an ML model $g_{\theta}$ (with parameters $\theta$) on the patients with $t_{i} = 0$ (the untreated patients) to predict the outcome based on their covariates. Specifically, $w_{i} = g_{\theta}(\boldsymbol{x}_{i})$ represents the probability that $y_{i} = 1$ or the probability that patient $i$ had a recurrence. This model will then be used in the $t = 1$ group to predict the baseline recurrence risk of these patients if they were not treated with imatinib; given that all patients in this group were treated with imatinib, these baseline recurrence risk predictions represent counterfactuals. 

Next, we split the patients into $m$ buckets based on their recurrence risk estimates $w_{i} = g_{\boldsymbol{\theta}}(\boldsymbol{x}_{i})$ for $i=1, \cdots, n$. Each bucket $k=1, \cdots, m$ is defined by an interval $[\underbar{$w$}_{k}, \overline{w}_{k}]$ of baseline recurrence risk estimates and each patient $i$ whose risk estimate $w_{i}$ is in $[\underbar{$w$}_{k}, \overline{w}_{k}]$ belongs to bucket $k$. Then, each patient is characterized by $(x_{i}, t_{i}, y_{i}, w_{i}, b_{i})$, where $b_{i}$ is the bucket of patient $i$. Note that $b_{i}$ is a function of $w_{i}$. The risk bucket thresholds and the number of patients in each bucket heavily depend on the problem and the data at hand.

Once we select the bucket thresholds, it is possible that the number of untreated patients is larger than the number of treated patients for a bucket $k$ or vice versa. To remedy this and make the cohort more similar to that of an RCT, we use an optimization algorithm to ensure that each patient in a bucket is equally likely to receive or not receive the treatment. More formally, consider the probability that a patient received the treatment $t_{a} \in \{0, 1\}$ given that they belong to bucket $k$, i.e., $p_{t|w}(t=t_{a} | \underbar{$w$}_{k} \leq w \leq \overline{w}_{k})$. In our approach, we require that the number of treated patients $n_{1}^{k}$ is equal to the number of the untreated patients $n_{0}^{k}$ in each bucket $k$. This is accomplished through a matching process and an optimization algorithm, which constitute a principled undersampling method. See next paragraph for a detailed description of the matching methodology. After matching, $p_{t|w}(t=0 \; \; | \; \; \underbar{$w$}_{k} \leq w \leq \overline{w}_{k}) = p_{t|w}(t=1 \; \; | \; \; \underbar{$w$}_{k} \leq w \leq \overline{w}_{k})$ for all buckets $k$. Simply put, the empirical probability that a patient received the treatment is the same as the empirical probability that the patient did not receive the treatment given that the patient belongs to bucket $k$. This emulates a randomized assignment of treatments within each risk stratum. 

To illustrate the matching process, suppose that there are fewer treated patients than untreated patients in bucket $k$. Our goal is to match each patient with imatinib to a patient without imatinib. The distance between patient $i$ and patient $j$ is $\| \boldsymbol{x}_{i} - \boldsymbol{x}_{j} \|$, which is the $\ell_{2}$ norm of the difference between the covariates. Usually, we normalize each covariate, which is common practice for distance metrics. The proposed optimization algorithm minimizes the total distance between all matched patients \cite{bertsimas2005optimization}. The use of mixed-integer optimization for matching observational data has also been proposed in other settings \cite{sun2016mutual, rosenbaum2020modern}. In this case, we also minimize the total distance between the selected patients. However, we match separately within each bucket or risk stratum. We use the following algorithm separately for each bucket, as required. Consider bucket $k$. $\mathcal{S}_{0}^{k}$ is the set of patients in bucket $k$ who did not receive imatinib, while $\mathcal{S}_{1}^{k}$ is the set of patients in bucket $k$ who received imatinib. The formulation is

\begin{equation}
    {\everymath{\displaystyle}
    \begin{array}{lll}
         \min_{\boldsymbol{z}} & \sum_{i \in \mathcal{S}_{1}^{k}}  \sum_{j \in \mathcal{S}_{0}^{k}} z_{ij} \| \boldsymbol{x}_{i} - \boldsymbol{x}_{j} \|^{2}  \\
         \text{s.t.} & \sum_{j \in \mathcal{S}_{0}^{k}} z_{ij} = 1, & \forall i \in \mathcal{S}_{1}^{k}, \\
         & \sum_{i \in \mathcal{S}_{1}^{k}} z_{ij} \leq 1, & \forall j \in \mathcal{S}_{0}^{k} \\
         & z_{ij} \in \{0, 1\}, & \forall i \in \mathcal{S}_{1}^{k}, \; \; j \in \mathcal{S}_{0}^{k}.
    \end{array}
    }
    \label{eq:eq1}
\end{equation}

The variable $z_{ij}=1$, if patient $i$ is matched to patient $j$; otherwise $z_{ij}=0$. The first constraint ensures that each patient who received imatinib is matched to a patient who did not receive imatinib. The second constraint ensures that each patient who did not receive imatinib is matched to no more than one patient who received imatinib. Once we obtain the solution to Problem $\eqref{eq:eq1}$, we only retain the $n_{s}$ matched patients in our final dataset.

This first step constructs strata of matched patients (untreated and treated) with similar baseline recurrence risk estimates within each stratum. This reduces prognostic heterogeneity and creates a cohort that emulates an RCT, as the patients in the $t = 0$ and $t = 1$ groups have similar baseline recurrence risk estimates. However, these baseline estimates are calculated using a model that was trained in the $t = 0$ group, and unobserved confounders are expected to be present in the $t_{i} = 1$ group. This means that the observed outcomes of the $t_{i} = 1$ group will be worse than their baseline recurrence risk estimates. Thus, we cannot attribute differences in outcomes between the matched groups to the treatment, as the $t = 1$ group may have worse outcomes due to the presence of unobserved confounding.

\subsection{Counterfactual Models and Reward Estimation}

In Step 2, we first quantify the impact of unobserved confounding on the estimates for the matched $t = 1$ group. To do this, we train two new 
classification models that estimate the risk of recurrence: one model, $h_{t=1}(\boldsymbol{x})$, in the matched $t = 1$ group and another, $h_{t=0}(\boldsymbol{x})$, in the matched $t = 0$ group. This is the \say{direct} reward estimation and is used to evaluate different policies or treatments \cite{dudik2011doubly}. It is worth noting that the training of both models in matched cohorts, with an identical number of patients in each stratum, enables them to \say{focus} on the same patient subgroups. We then compare the mean recurrence risk estimates $\hat{w}_{t=0} = \frac{1}{n_{s}} \sum_{i=1}^{n_{s}} h_{t=0}(\boldsymbol{x}_{i})$ and $\hat{w}_{t=1} = \frac{1}{n_{s}} \sum_{i=1}^{n_{s}} h_{t=1}(\boldsymbol{x}_{i})$ for the two treatment options across all $n_{s}$ selected patients. If $\hat{w}_{t=1} > \hat{w}_{t=0}$ this is evidence of unobserved confounding in the treated group, and the distance between the mean recurrence risk estimates or $\hat{w}_{t=1} - \hat{w}_{t=0}$ is the minimum amount of unobserved confounding bias. Note that we say minimum, because it is unknown whether more unobserved confounding bias is present. The solution to this issue is to increase the weight of patients from the treated group who did not have the outcome (i.e., recurrence) as this will decrease the mean recurrence risk estimates for the entire treated group.  The ultimate goal is to achieve $\hat{w}_{t=1} = \hat{w}_{t=0}$. 

More formally, we train  the classification models $h_{t=0}(\boldsymbol{x})$ and $h_{t=1}(\boldsymbol{x})$ by minimizing a loss function $\ell$, such as the cross-entropy loss. Let $\mathcal{S}_{1}$, $\mathcal{S}_{0}$ be the set of the $n_{1}^{s}$, $n_{0}^{s}$ selected patients who received or did not receive imatinib, respectively. In addition, let $\mathcal{S}_{1}^{-}$, $\mathcal{S}_{1}^{+}$ be  the set of patients in $\mathcal{S}_{1}$ who had or did not have a recurrence, respectively.
 In order to train the classification model $h_{t=0}(\boldsymbol{x})$ we use the empirical loss 
 \begin{equation}
 \label{0training}
    {\everymath{\displaystyle}
        \frac{1}{n_{0}^{s}} \sum_{i \in \mathcal{S}_{0}} \ell(y_{i}, h_{t=0}(\boldsymbol{x}_{i})).
    }
\end{equation}
In order to train the classification model $h_{t=1}(\boldsymbol{x})$ we use the empirical loss 
\begin{equation}
 \label{1trainng}
    {\everymath{\displaystyle}
        \frac{1}{n_{1}^{s}} \left(\sum_{i \in \mathcal{S}_{1}^{-}} \ell(y_{i}, h_{t=1}(\boldsymbol{x}_{i})) + \rho \sum_{j \in \mathcal{S}_{1}^{+}} \ell(y_{j}, h_{t=1}(\boldsymbol{x}_{j}))\right).
    }
\end{equation}
Suppose we increase the weight $\rho$ for the patients who received imatinib but did not have a recurrence.
The model will then try to make better predictions for the patients with the larger weight, since this yields a greater improvement in the loss \cite{elkan2001foundations}. This approach is called cost-sensitive learning and is very common in settings with imbalanced data \cite{krawczyk2016learning}. In this case, the data are not necessarily imbalanced and the weight parameter is used to bias the model towards treated patients with \say{good} outcomes. To the best of our knowledge, this is the first work in which cost-sensitive learning is utilized in this manner. 

Using the weighted loss, the risk recurrence estimates of $h_{t=1}(\boldsymbol{x})$ also depend on the weight $\rho$. To make this explicit, we use the  notation $h_{t=1}(\boldsymbol{x},\rho)$. In this case, we expect $\frac{1}{n_{s}} \sum_{i=1}^{n_{s}} h_{t=1}(\boldsymbol{x}_{i},\rho)$ to increase with $\rho \geq 1$. Note   $h_{t=0}(\boldsymbol{x})$ does not depend on $\rho$ as the training function \eqref{0training} does not depend on $\rho$.

By adjusting the weight so that $\hat{w}_{t=1} = \hat{w}_{t=0}$, we eliminate the effect of the minimum unobserved confounding. However, there may be residual unobserved confounding, meaning that $\hat{w}_{t=1}$ is still greater than the real mean risk of recurrence under the treatment. To address this issue, we also provide a principled method for finding the optimal weight based either on a held-out validation set or on cross-validation. The method is described in detail in Section \ref{sec:val}.

\subsection{OPTs}
\label{sec:opts}

Applying the first two steps will result in a cohort of patients with similar baseline prognosis within each stratum and with greatly reduced unobserved confounding, emulating an RCT cohort. We can now apply a methodology to define the subgroups of patients with distinct characteristics who have HTE. To do so, we will use the probabilities derived from the two counterfactual models we previously trained to train an OPT with direct reward estimation, which can group patients with similar HTE in each of its nodes or leaves \cite{amram2022optimal}. The OPT solves the following problem using a tree-based model

\begin{equation}
    {\everymath{\displaystyle}
        \min_{\tau(\cdot)} \sum_{i=1}^{n_{s}} \left(1_{\{\tau(\boldsymbol{x}_{i})=0\}} h_{t=0}(\boldsymbol{x}_{i}) + 1_{\{\tau(\boldsymbol{x}_{i})=1\}} h_{ t=1}(\boldsymbol{x}_{i},\rho) \right),
    }
\end{equation}
where $\tau(\boldsymbol{x})$ is a policy that assigns treatments to patients based on their features $\boldsymbol{x}$ and $1_{\{ \cdot \}}$ is the indicator function that takes value 1 if its argument is true, and 0,
otherwise. 

Considering that the OPT's objective is to minimize the risk of recurrence across all patients, and the fact that weight increase in Step 2 resulted in a reduced risk of recurrence for those who received imatinib, it can be expected that OPT will tend to favor the assignment of imatinib. \

Of note, HTE depends both on the baseline risk of the outcome and the response to the treatment. As the baseline recurrence risk increases, a greater absolute recurrence risk reduction is expected. For example, the same relative risk reduction (RRR) of $50\%$ will result in an absolute risk reduction (ARR) of $40\%$ if the baseline risk is $80\%$ but only $20\%$ if the baseline risk is $40\%$. Thus, we can compare the ARR and RRR within each node of the OPT. Specifically, the ARR can be used to decide which patients should be offered a treatment depending on what risk reduction threshold is deemed clinically meaningful. The RRR can indicate whether the response to the treatment varies according to specific patient characteristics. 

\begin{algorithm}[]
	\caption{The R.O.A.D. to precision medicine. An example of observational data of patients with GIST with treatment and no treatment options.} 
	\label{alg:alg_1}
	\SetKwComment{Comment}{/* }{*/}
	\SetAlgoLined 
	\KwIn{\text{Original dataset} ($\boldsymbol{X}$, $\boldsymbol{y}$) \\
	\textbf{Parameters:} \SetKwData{}{} \\
	\begin{itemize}
	    \item $m$: number of risk buckets
	    \item $[\underbar{$w$}_{k}, \overline{w}_{k}]$: risk bucket thresholds for $k=1, \cdots, m$
	    \item $g_{\theta}$: ML model for predicting the baseline risk of each patient (default: RF)
            \item $h_{t=0}, \; \; h_{t=1}$: counterfactual models for predicting the outcome (risk of recurrence) under each treatment (default: RF) 
            \item $\rho$: weight of the patient group of interest (treated with no recurrence)
	    \item $OPT$: Optimal Policy Tree that optimally assigns        treatments to subgroups of patients
	\end{itemize}} 
	\KwOut{Trained $OPT$}
    \Comment{Step 1: Risk Stratification and Matching}
    Train $g_{\theta}$ on the untreated patients in $(\boldsymbol{X}, \boldsymbol{y})$ \\
    Split data on $m$ datasets $(\boldsymbol{X}^{k}, \boldsymbol{y}^{k})$ based on $g_{\theta}$ and $[\underbar{$w$}_{k}, \overline{w}_{k}]$ for $k=1, \cdots, m$  \\
    \For{$k=1, \cdots, m$}{
        $n_{0}^{k}, n_{1}^{k} = $ the number of untreated, treated patients in bucket $k$ \\
        \If{$n_{0}^{k} \; \; != \; \; n_{1}^{k}$}{
        \Comment{Select the patients using the proposed matching algorithm}
            $(\boldsymbol{X}^{k}, \boldsymbol{y}^{k}) = sel (\boldsymbol{X}^{k}, \boldsymbol{y}^{k})$   
        }
    }
    $(\boldsymbol{X}, \boldsymbol{y}) = $ concatenate datasets $(\boldsymbol{X}^{k}, \boldsymbol{y}^{k})$ for $k=1, \cdots, m$ \\
    $(\boldsymbol{X}_{0}, \boldsymbol{y}_{0}), (\boldsymbol{X}_{1}, \boldsymbol{y}_{1}) = $ the untreated, treated patients in $(\boldsymbol{X}, \boldsymbol{y})$ \\
    \Comment{Step 2: Cost-sensitive Counterfactual Estimation}
    Train $h_{t=0}$ on $(\boldsymbol{X}_{0}, \boldsymbol{y}_{0})$ \\
    Train $h_{t=1}(\cdot, \rho)$ on $(\boldsymbol{X}_{1}, \boldsymbol{y}_{1})$ with $\rho=1$ \\
    $\hat{w}_{t=0} = \frac{1}{n_{s}} \sum_{i=1}^{n_{s}} h_{t=0}(\boldsymbol{x}_{i})$, the mean rewards under $t=0$ \\
    $\hat{w}_{t=1} = \frac{1}{n_{s}} \sum_{i=1}^{n_{s}} h_{t=1}(\boldsymbol{x}_{i}, \rho)$, the mean rewards under $t=1$ \\
    \Comment{Gradually increase the weight unti treatment is more effective than no treatment}
    $\epsilon = 0.1$ \\
    \While{$\hat{w}_{t=0} < \hat{w}_{t=1}$}{
        $\rho = \rho + \epsilon$ \\
        Train $h_{t=1}(\cdot, \rho)$ on $(\boldsymbol{X}_{1}, \boldsymbol{y}_{1})$ with $\rho=1$ \\
        $\hat{w}_{t=1} = \frac{1}{n_{s}} \sum_{i=1}^{n_{s}} h_{t=1}(\boldsymbol{x}_{i})$
    }
    \Comment{Step 3: Train OPT}
    Train $OPT$ on the counterfactual estimations $(\boldsymbol{x}_{i}, h_{t=0}(\boldsymbol{x}_{i}))$ and $(\boldsymbol{x}_{i}, h_{t=1}(\boldsymbol{x}_{i}, \rho))$ for all $\boldsymbol{x}_{i}$ in $\boldsymbol{X}$ \\
    \Comment{Step 4: Validation}
    Validate the $OPT$ using the sensitivity and specificity metrics using cross-validation or a held-out validation set. \\
    \Comment{Step 5: Weight Tuning}
    Repeat Steps 2-4 to find the optimal weight $\rho$ by increasing $\rho$ at each iteration \\
    \Return $OPT$

\end{algorithm} 

\subsection{Validation}
\label{sec:val}

In this step,  we want to validate our findings in external cohorts that did not receive the treatment in question and thus their baseline risks are not modified. To do so, we employ the well-known statistical concepts of sensitivity and specificity, which is the first time such concepts have been applied to the HTE field. We selected sensitivity and specificity because neither are influenced by the prevalence of an outcome (e.g., recurrence) and thus are robust across different populations. Furthermore, these metrics can be used to quantify the validity of the HTE recommendations as they take numerical values and are interpretable. In our example, the sensitivity of the OPT recommendations is assessed by calculating the ratio of patients who recurred but for whom imatinib was recommended by the OPT, divided by the sum of these patients and those who eventually recurred but for whom the OPT failed to recommend imatinib. For example, a sensitivity of $90\%$ means that of 100 untreated patients who recurred and thus would have benefited from treatment, the OPT managed to correctly assign a treatment to 90 patients. Of note, this assumes that imatinib is an effective drug that would have prevented the recurrences. The specificity of the OPT recommendations is evaluated by calculating the ratio of patients who did not recur for whom imatinib was not recommended by the OPT, divided by the sum of these patients and those who did not recur but for whom the OPT recommended imatinib. For example, a specificity of $80\%$ means that of 100 untreated patients who did not recur and thus should not have been treated with imatinib, the OPT managed to spare 80 patients from unnecessary treatment. Finally, given that tuning the hyperparameters of any decision tree can generate trees with different structures, we propose the use of sensitivity and specificity to assess the robustness of the tree recommendations. Specifically, if the sensitivity and specificity of several OPTs converge, this indicates a robustness of the method regardless of which OPT is selected. This is important as the selection of a single tree can be very subjective. Our algorithm is also provided in Algorithm \ref{alg:alg_1}. 

\paragraph{Weight Tuning} In our method, we also use sensitivity and specificity to tune the weight $ \rho$ of the loss function of the counterfactual models. First, we split the data into a training set and a validation set. We then use the training set to train models with different weights and the validation set to select the weight with the best combination of sensitivity and specificity. Note that this is a repurposing of the validation and cross-validation approaches that are widely used to tune the hyperparameters of ML models. 

If we want to reduce overtreatment, we seek to increase the specificity while maintaining high sensitivity. Similarly, if we want to reduce undertreatment, we seek  to increase the sensitivity. This indicates which patient groups will be assigned increased weight. Specifically, if we want to increase sensitivity, we will increase the weight of the patients who received the treatment and did not have a recurrence. As detailed in Section \ref{sec:opts}, this will result in OPT assigning imatinib to a larger proportion of patients. When more patients are assigned imatinib, a greater number of recurrences will be prevented by imatinib. This is the reason for the observed increase in sensitivity. On the contrary, if we want to increase specificity, we will increase the weight of the patients who did not receive the treatment and did not have a recurrence. For instance, in the GIST example, we test different weights ranging from 1 to 3 for the patients who received imatinib and did not experience recurrence. Of note, sensitivity is generally the more important metric in the medical field, as we do not want to miss patients who would have benefited from treatment, thereby complying with the Hippocratic principle of \say{first, do no harm.}

\section{Results}

The results section is divided into two components. In the first segment, we will employ a clinical case to illustrate how our framework can be applied to analyze observational data. In the second portion, we will utilize a different clinical example to showcase the effectiveness of our framework in analyzing randomized clinical trial data.

\subsection{Demonstration of the proposed framework in an observational cohort}
\noindent 
\textit{Clinical problem}

\noindent The evolution of GIST treatment reflects the success of targeted therapies in solid tumors, as imatinib has dramatically improved survival in patients with advanced GIST \cite{corless2011gastrointestinal}. In fact, the median overall survival has improved from 18 months to more than 70 months. Although the efficacy of adjuvant imatinib has been proven for patients with localized GIST, optimal patient selection remains controversial despite multiple RCTs \cite{dematteo2009adjuvant, heinrich2017defining, blay2021gastrointestinal}.

Identifying which patients benefit from adjuvant imatinib is clinically important so as to balance between undertreating those who would benefit and overtreating those who have already achieved cure from surgical resection. Of note, current selection criteria are liberal and any patient with at least intermediate risk of recurrence is considered for therapy \cite{casali2015time}. However, it remains unclear what threshold of “intermediate” recurrence risk should be used to justify adjuvant imatinib. This uncertainty is reflected in the guidelines from the European Society of Medical Oncology, which currently recommends “adjuvant imatinib therapy for patients with a significant risk of relapse, with room for shared decision-making when the risk is intermediate” \cite{casali2018gastrointestinal}. One explanation for this uncertainty is the lack of a placebo arm with intermediate risk patients in most RCTs given ethical concerns.

\noindent\textit{Original cohort}\\
\noindent The original cohort included observational data from 536 patients who underwent surgery at Memorial Sloan Kettering (MSK) between 1982 and 2017. The characteristics of these patients are presented in Table \ref{tab1}. At a median follow-up of 87 months (IQR: 54–126), 100 of 536 patients (18.6\%) developed recurrence, with a 7-year recurrence free survival (RFS) rate of 80\% (95\%CI: 77-84\%). A total of 117 patients received imatinib while 419 did not. The median follow-up for those who received imatinib was 69 months (IQR: 39-103) and the 7-year RFS rate was 57\% (95\%CI: 47-70\%). The median follow-up for those who did not receive imatinib was 91 months (IQR: 59-134) and the 7-year RFS rate was 86\% (95\%CI: 82-89\%). The Kaplan Meier (KM) plot for these two groups is shown in Figure 1A. Notably, the log rank test showed that RFS was significantly worse in patients who received imatinib (p $< 0.001$); since we know from RCT data that imatinib generally benefits patients, this finding indicates the presence of confounding.

\begin{table}[hbtp]
\begin{threeparttable}
\caption{Clinicopathological and treatment characteristics of patients with GIST.}
\label{tab1}
\small
\begin{tabularx}{\textwidth}{|l||X|X|X|}
\hline
 \textbf{Characteristics} & \textbf{MSK cohort} & \textbf{Polish registry dataset} & \textbf{P-value}\\ \hline
\textbf{Age(median [IQR])} &	65[54,72]	& 60[52,68]	& $<$0.001\\ \hline
\textbf{Sex(\%)} &	& &	0.069\\ \hline
   Female &	267(49.8)	& 308(55.3) &	\\ \hline
   Male	 & 269(50.2) &	249(44.7) &	\\ \hline
\textbf{Max tumor size(median [IQR])} &	4.5[3,8] &	6[4,9.125]&	$<$0.001\\ \hline
\textbf{Mitotic count numeric(median [IQR])} &	3[1,8]&	3[1,9]&	0.070\\ \hline
\textbf{Site categorical(\%)} &		& &	$<$0.001\\ \hline
   Gastric	& 391(73)	& 320(57.5) &	\\ \hline
   Non gastric &	145(27)	& 237(42.5)&	\\ \hline
\end{tabularx}
\end{threeparttable}
\end{table}

A random forests (RF) classifier was trained in the patients who did not receive imatinib to explore the relationship between the three most established prognostic factors (i.e., tumor size, mitotic count, and tumor site) and the baseline risk of recurrence (i.e., if no adjuvant treatment was administered after surgery). The same classifier was then used to calculate the counterfactual baseline risk of recurrence of the patients who were treated with imatinib; in other words, we calculated what the risk of recurrence would be if the patient had not received imatinib. The model predictions were then used to stratify the patients by their baseline risk of recurrence. The prognostic strata are presented in Figure 2A, which demonstrates the considerable imbalance in the distribution of baseline recurrence risk between patients who received and did not receive imatinib. The presence of confounding, which was suspected upon visual inspection of the KM plots, was confirmed as most patients who did not receive imatinib had a low baseline risk of recurrence while most patients who received imatinib had a higher baseline risk of recurrence.

\begin{figure}[hbtp]
\centering
\caption{Recurrence-free survival after GIST resection stratified by receipt of adjuvant imatinib before (A) and after (B) prognostic matching.}
\label{figB}
\includegraphics[width=1\linewidth]{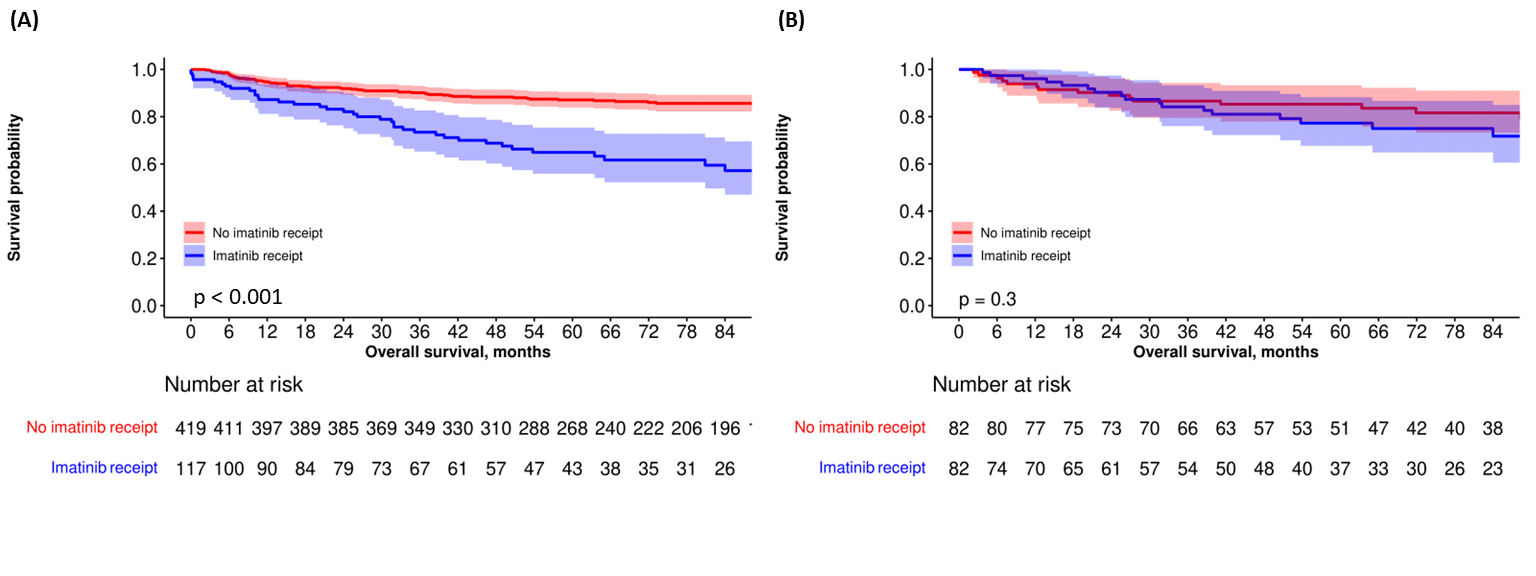}
\end{figure}

\begin{figure}[hbtp]
\centering
\caption{Distribution of patients with GIST according to their baseline risk of recurrence before (A) and after (B) prognostic matching.}
\label{figC}
\includegraphics[width=0.85\linewidth]{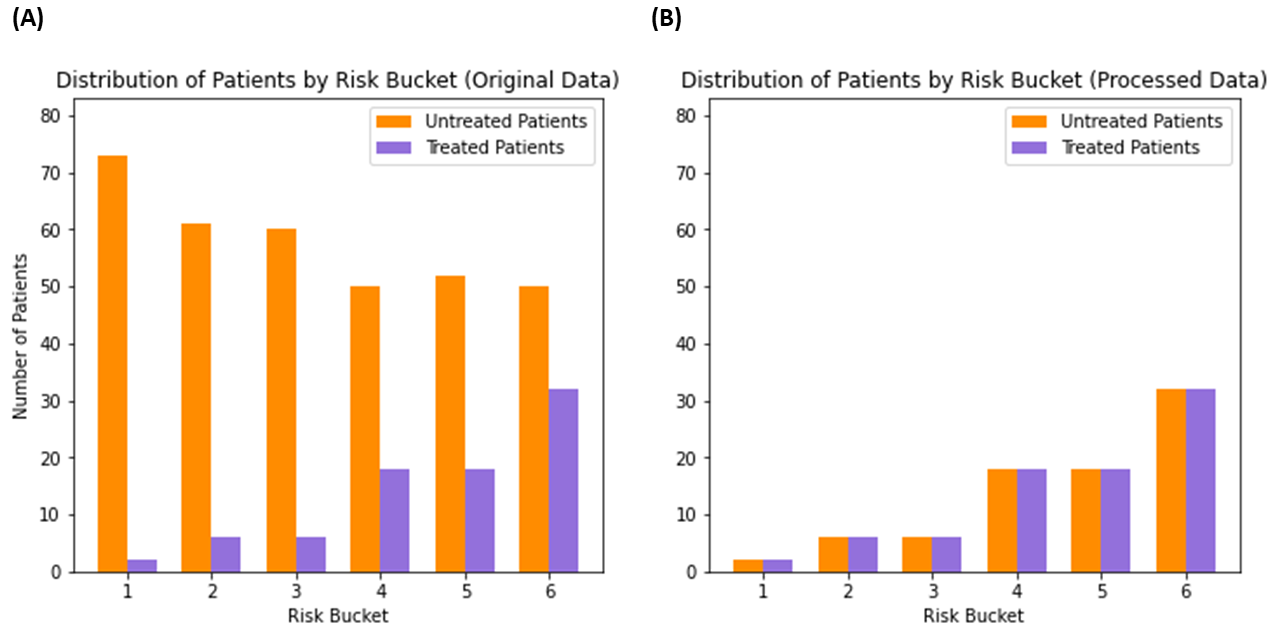}
\end{figure}

\noindent\textit{Matched cohort}\\
\noindent Prognostic matching created a cohort of 164 patients, of which 82 received imatinib and 82 did not. Figure \ref{figC}B demonstrates that prognostic matching remedied the previously observed imbalance in baseline recurrence risk between the treated and untreated patients. In the matched cohort, the median follow-up for patients who received imatinib was 71 months (IQR: 44-103), and the 7-year RFS rate was 72\% (95\%CI: 61-85\%). The median follow-up for patients who did not receive imatinib was 93 months (IQR: 57-145), and the 7-year RFS rate was 82\% (95\%CI: 73-91\%). The difference in the 7-year RFS between the treated versus untreated patients decreased after prognostic matching (Figure \ref{figB}B).

This mostly stems from an increase in the 7-year RFS rate of the matched patients who received imatinib, as the matched patients who did not receive imatinib had only a slight decrease in their RFS rates. This is important as we want to capture the average effect of the treatment and not the treatment effect of the treated (TTE). However, the matched patients who received imatinib still had a lower 7-year RFS rate than the matched untreated patients (p = 0.3), which indicates the presence of residual unobserved confounding.

\begin{figure}[hbtp]
\centering
\caption{Optimal policy trees  developed with a  weight of $\rho=1.5$.}
\label{fig1}
\includegraphics[width=1\linewidth]{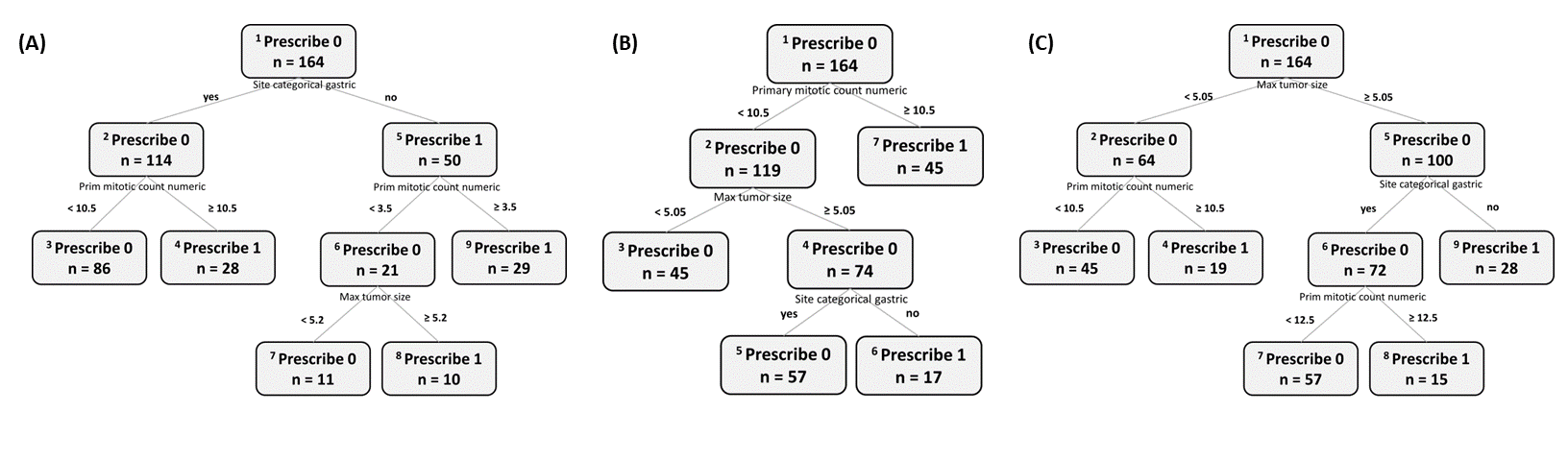}
\end{figure}

To address the residual unobserved confounding, we first trained a survival RF model in the matched treated patients and a separate survival RF model in the matched untreated patients. We then experimented by serially increasing the weight $\rho$ for patients who received imatinib and did not experience a recurrence; the weights we used ranged from 1.5 to 3.25. Although the predicted 7-year RFS under imatinib improved from 77\% to 79\% with a weight of 1.5, the effect of unobserved confounding was still evident as the predicted 7-year RFS was still lower under imatinib versus no imatinib (79\% vs 82\%). In contrast, when we used a weight of 2, predicted 7-year RFS was higher under imatinib versus no imatinib (83\% vs 82\%). This means that a weight of at least 2 is needed to address the effects of residual unobserved confounding. To find the \textit{optimal} weight (i.e., that with the best combination of sensitivity and specificity with a greater emphasis on optimizing sensitivity), we used weight as a “hyperparameter” and tuned it by measuring sensitivity and specificity in a validation cohort, which is described below in the validation section. As demonstrated in Table 2, sensitivity rises with greater weight, while specificity decreases in accordance with the familiar trade-off. Of note, the recommendations of the various OPTs that were trained using the same weight were robust with regard to their sensitivity and specificity. For example, sensitivity varied from 82 to 85\% and specificity varied from 76 to 79\% across all OPTs that were trained by tuning the hyperparameters with the minimum weight of 1.5. Similarly, sensitivity varied from 91 to 94\% and specificity varied from 52 to 55\% across all OPTs that were trained by tuning the hyperparameters with the maximum weight of 3.25. The robustness of the OPTs trained through hyperparameter tuning with the same weight is also reflected in their tree structures. To illustrate this, we present three examples of OPTs in Figure \ref{fig1}, all three developed with  $\rho=1.5$. As shown in the figure, it is clear that the cutoff values for mitotic count and tumor size are very similar. 

We ultimately opted for a weight of 2.25 due to the fact that, with greater weights, sensitivity improved only marginally, while specificity declined disproportionately. Specifically, although we chose to increase the weight from 1.5 to 2.25 and sacrifice 6\% of specificity in order to gain 4\% in sensitivity, a further increase of the weight from 2.25 to 3.25 that would sacrifice another 18\% of specificity for a modest gain of only 5\% in sensitivity (Table \ref{tab2}) was deemed unacceptable by us.  The weight of 2.25 led to a predicted 7-year RFS under imatinib that surpassed the rate under no imatinib (85\% vs 82\%), and it also allowed to train an OPT with a sensitivity and specificity of 89\% (95\%CI: 84-93\%) and 70\% (95\%CI: 65-74\%), respectively. On a broader note, depending on the specific characteristics of the problem, the end user has the flexibility to determine which of the two metrics to prioritize.

\begin{table}[hbtp]
\caption{The correlation between weight tuning and OPT sensitivity and specificity metrics.}
\label{tab2}
\centering
%\small
\setstretch{1}
\begin{tabular}{|c||c|}
\hline
 \textbf{Weight} & \textbf{Sensitivity and Specificity of OPT} \\ \hline
1.50  &    85 and 76\% \\ \hline
2.00    & 84 and 77\%    \\ \hline
\textbf{2.25}  &   \textbf{89 and 70\%} \\ \hline
2.85   &   93 and 55\% \\ \hline
3.05  &  93 and 53\% \\ \hline
3.25 & 94 and 52\% \\ \hline
\end{tabular}
\end{table}

\begin{figure}[hbtp]
\centering
\caption{Optimal Policy Tree for recurrence-free survival in patients with GIST.}
\label{fig3}
\includegraphics[width=0.7\linewidth]{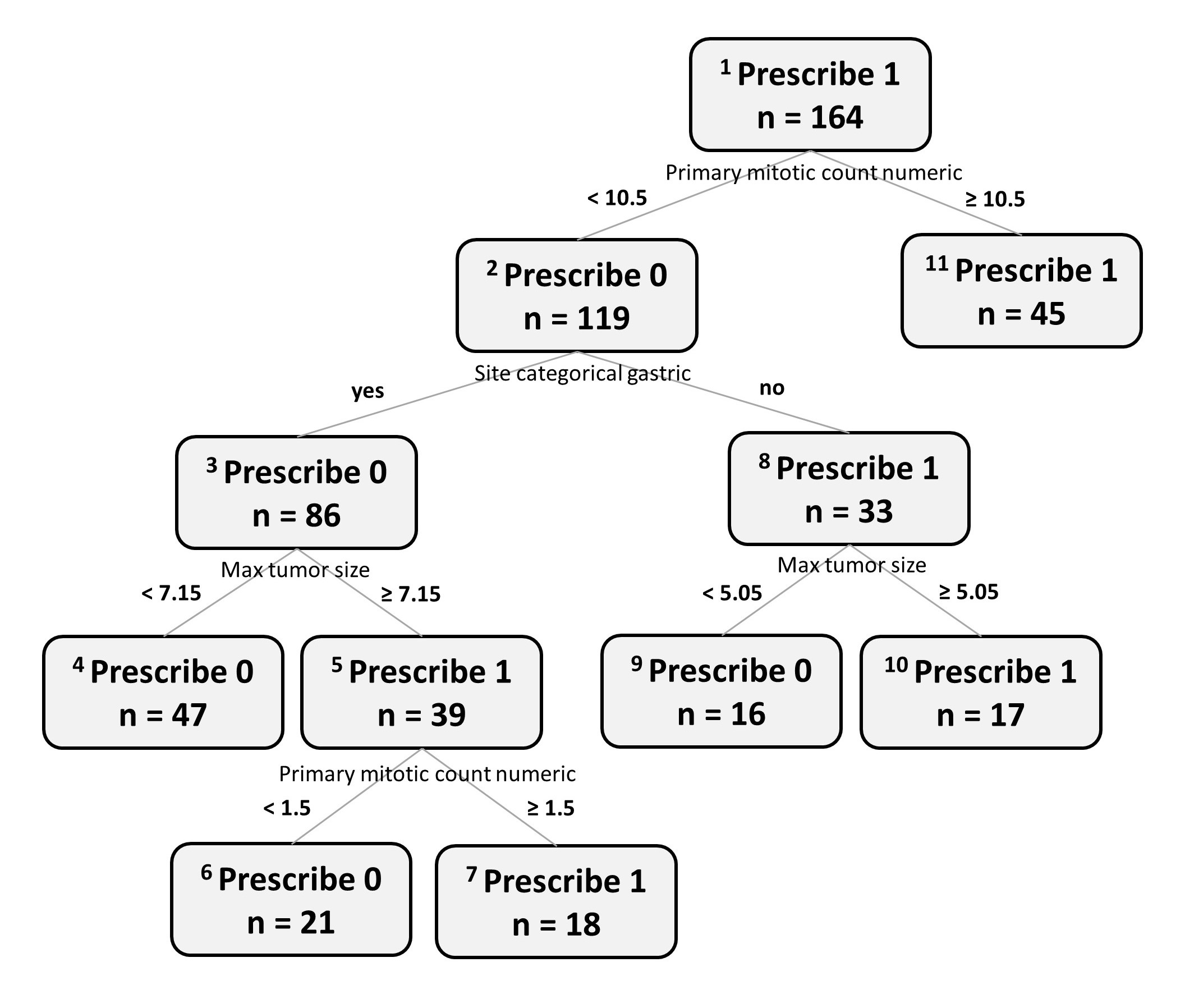}
\end{figure}

\noindent\textit{OPT}\\
\noindent The OPT that was trained using the optimal weight is illustrated in Figure \ref{fig3}.  Its hyperparameters were a minbucket of 15 and a max depth of 4. The OPT recommended imatinib for three patient subgroups and no adjuvant treatment for the other three subgroups. The subgroups for which treatment was not recommended included patients with a mitotic count less than 10.5 and small tumors of either non-gastric ($< 5.05$ cm) or gastric ($< 7.15$ cm) origin. This recommendation did not change for patients with larger tumors ($> 7.15$ cm) as long as the mitotic count was low ($< 1.5$) and the site was gastric. In contrast, adjuvant imatinib was recommended for patients with a mitotic count equal or greater than 10.5 and for those with tumors of non-gastric sites that were equal or larger than 5.05 cm. It was also recommended for patients with tumors of large size ($> 7.15$ cm), gastric site, and a mitotic count equal or greater than 1.5.

\noindent\textit{Validation cohort}\\
\noindent The validation cohort included 557 patients from Poland. The characteristics of these patients are presented in Table 1.  At a median follow-up of 61 months (IQR: 29–94), 190 of 557 patients (34\%) developed recurrence. The 7-year RFS rate was 54\% (95\%CI: 49-60\%), which is worse than that of untreated patients in the training cohort. This is expected because many patients in the training cohort who did not receive imatinib were treated in an era when imatinib was available, which means that they likely did not receive it because they were deemed to have a low baseline recurrence risk. In contrast, the validation cohort consists of patients treated in the pre-imatinib era; thus, the baseline recurrence risks of these patients represent the entire prognostic spectrum of the disease.

The sensitivity and specificity of the OPT recommendations were 89\% (95\%CI: 84-93\%) and 70\% (95\%CI: 65-74\%), respectively. Since this cohort is from the pre-imatinib era and thus reflects the natural course of the disease, we can use the prevalence of recurrence and calculate the negative predictive value (NPV) of an OPT recommendation that imatinib should not be offered to a patient. Of note, the NPV was 93\% which is the likelihood that an individual who does not receive imatinib per OPT recommendations will not recur. Importantly, the sensitivity of the OPT recommendations surpassed that of the criteria used to assess eligibility in the SSG XVIII/AIO randomized trial (89\% vs 82\%) \cite{joensuu2012one}. The OPT recommendations were also compared to the criteria used to assess eligibility in the PERSIST-5 clinical trial \cite{raut2018efficacy}. Although they had similar sensitivity (OPT: 89\%; PERSIST-5: 91\%), the specificity of the OPT recommendations was higher (70\% vs 63\%).

Because certain tumor characteristics have already been associated with a lower risk of recurrence, it is important that the OPT recommendations reflect these established relationships. For example, a non-gastric tumor site is a known risk factor for recurrence; the OPT recommendation was consistent in that the tumor size threshold for offering imatinib was lower for non-gastric versus gastric tumors (5 vs 7 cm). In a similar example, the OPT recommended that patients with a gastric GIST greater than 7 cm and with a very low mitotic count should not receive imatinib, but recommended that those with a non-gastric GIST greater than 5 cm should receive imatinib regardless of the mitotic count. Of note, a mitotic count cut off of 10.5 was the first variable used by the OPT to define a subgroup of patients who should receive imatinib regardless of their other characteristics. Importantly, not only is mitotic count considered the strongest factor associated with a high risk of recurrence, but this specific cut off of 10.5 is also the same one used in the one-variable-at-a-time sub-analyses of the RCTs on imatinib. In fact, Joensuu and colleagues performed 7 subgroup analyses using one patient characteristic at a time and found that a mitotic count greater than 10 was the only characteristic significantly associated with a benefit from longer treatment with imatinib \cite{joensuu2020survival}. Although this was remarkably consistent with the OPT recommendation, it also highlights the fundamental limitation of the one-variable-at-a-time sub analysis. Specifically, patients with a mitotic count of less than 10.5 will inevitably differ by other tumor characteristics such as tumor size and site. Thus, serially dividing patients using this single characteristic (mitotic count $< 10.5$  vs $\geq 10.5$) may be grossly misleading. Indeed, as our OPT shows, patients with a mitotic count less than 10.5 are not a uniform cohort with regard to benefit from imatinib; we found that three patient subsets do not benefit from imatinib whereas two subgroups do.

\subsection{Demonstration of the proposed framework in a randomized trial cohort}
\noindent\textit{Clinical problem }\\
\noindent The benefit of adjuvant radiotherapy (RT) for patients with extremity or truncal sarcomas has been demonstrated in two prospective randomized trials; both showed that RT reduces the local recurrence rate for patients treated with limb-sparing surgery \cite{pisters1996long, yang1998randomized}. In turn, RT is currently administered to all patients who undergo limb sparing surgery despite some observational studies reporting low local recurrence rates without RT. Thus, it is unlikely that all these patients need treatment. In fact, one of these studies concluded that “further study is needed to carefully define this subset of patients,” but this has not been methodologically possible so far \cite{baldini1999long}.

\noindent\textit{Original cohort}\\
In this example, we use RCT data to demonstrate that our framework may solve this issue. The training cohort included 164 patients from a randomized trial conducted at MSK with updated follow up \cite{pisters1996long}. The characteristics of these patients are presented in Table \ref{tab3}; at a median follow-up of 176 months (IQR: 65-249), 38 of 164 patients (23\%) developed local recurrence. The 5-year local RFS rate was 77\% (95\%CI: 70-84\%).

\begin{table}[hbtp]
\begin{threeparttable}
\caption{Clinicopathological and treatment characteristics of patients with extremity sarcomas.}
\label{tab3}
\small 
\begin{tabularx}{\textwidth}{|l||X|X|X|}
\hline
\textbf{Characteristics} & \textbf{RCT cohort} & \textbf{MSK cohort} & \textbf{P-value}\\ \hline
\textbf{Age(median [IQR])} & 53[39,65] & 55[42-67] & 0.060\\ \hline
\textbf{Sex(\%)} & & &0.362\\ \hline
   Female & 70(43) & 438(47) &	\\ \hline
   Male & 93(57) & 498(53) &	\\ \hline
\textbf{Liposarcoma(\%)} & & &			$<$0.001\\ \hline
   Yes	& 10(6) & 370(40) &	\\ \hline
   No	& 154(94) & 566(60) &	\\ \hline
\textbf{Margin(\%)} & & 	&	0.338\\ \hline
   Positive &	28(17) & 133(14) &	\\ \hline
   Negative &	136(83)	& 803(86) &	\\ \hline
\textbf{Tumor grade(\%)} & & &	$<$0.001\\ \hline
   High	& 114(70) &	403(43)	 & \\ \hline
   Low	&  49(30)&	533(57) &	\\ \hline
\textbf{Specific tumor size(median [IQR])} &	7[4.9,10] &	6[3,12.4] &	0.085\\ \hline
\textbf{Depth description(\%)} & & &	0.076\\ \hline
   Deep	& 115(70) &	589(63) &\\ \hline	
   Superficial & 49(30) & 347(37) &	\\ \hline
\textbf{Site(\%)} & & &			$<$0.001\\ \hline
   Upper &	34(21) & 227(24) &	\\ \hline
   Lower &	104(63) & 643(69) &	\\ \hline
   Trunk &	26(16) & 66(7) &	\\ \hline
\textbf{Postoperative chemotherapy(\%)} & & & 	$<$0.001\\ \hline
   Yes & 63(39)	 & 0(0) &	\\ \hline
   No &	100(61)	 & 936(100) &	\\ \hline
\end{tabularx}
\end{threeparttable}
\end{table}

\begin{figure}[hbtp]
\centering
\caption{Local Recurrence-free survival after extremity soft tissue sarcoma resection stratified by receipt of adjuvant radiotherapy.}
\label{fig4}
\includegraphics[width=0.8\linewidth]{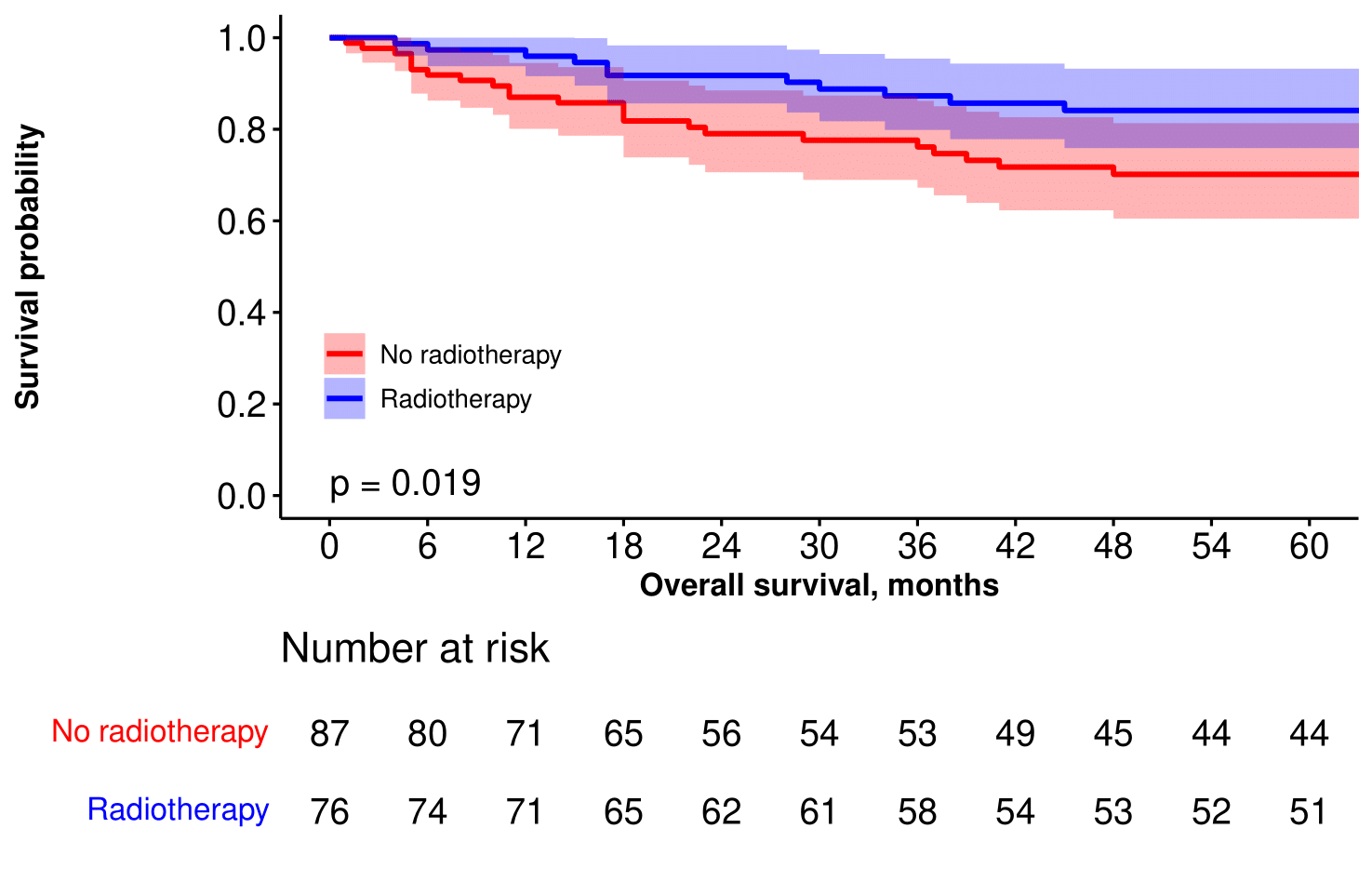}

\end{figure}

In this cohort, 77 patients received a form of RT called brachytherapy and 87 did not. The baseline local recurrence risk for all patients (including those who were treated with radiotherapy) was estimated using the MSK nomogram for extremity/truncal sarcomas, which is an established prognostic model \cite{cahlon2012postoperative}. The median follow-up for those who received RT was 183 months (IQR: 94-237), and the 5-year local RFS rate was 84\% (95\%CI: 76-93\%). The median follow-up for those who did not receive RT was 170 months (IQR: 45-283), and the 5-year local RFS rate was 70\% (95\%CI: 61-81\%). The KM plot for these two groups is shown in Figure \ref{fig4}. The log rank test showed that local RFS was significantly better in patients who received radiotherapy (p = 0.019); in contrast to our prior example, this result does not indicate confounding, which is expected since this is a randomized trial cohort.

\begin{figure}[hbtp]
\centering
\caption{Distribution of patients with extremity soft tissue sarcoma according to their baseline risk of local recurrence before (A) and after (B) oversampling of the higher risk strata.}
\label{A}
\includegraphics[width=0.85\linewidth]{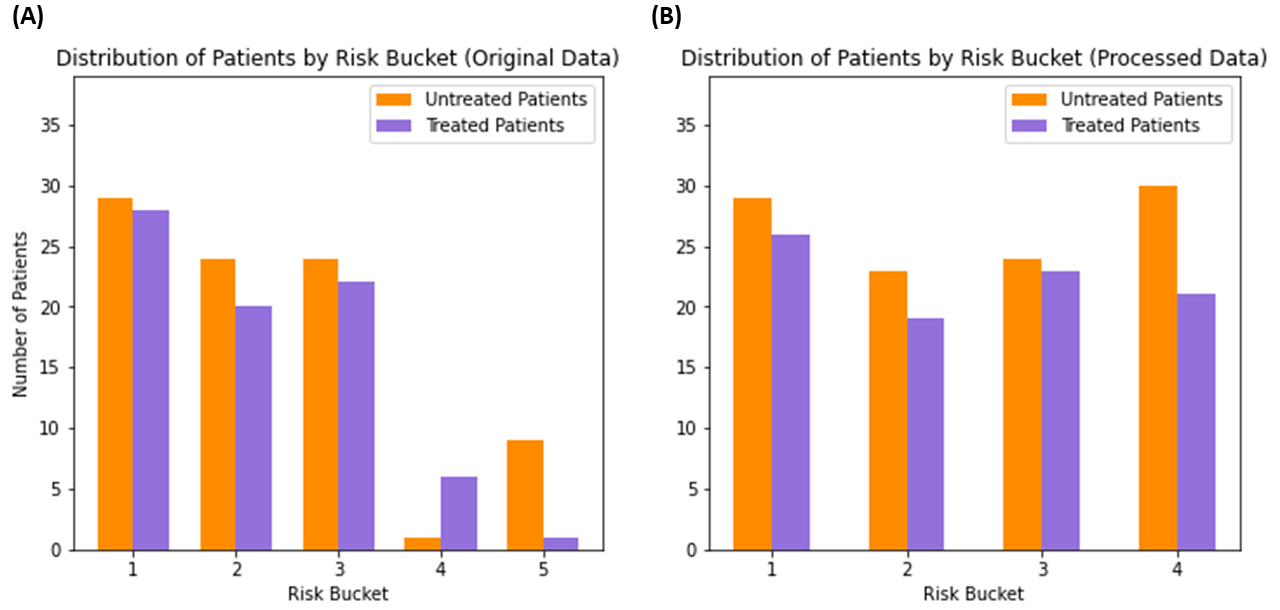}
\end{figure}

Patients were stratified into five prognostic strata according to their baseline risk of local recurrence (0-10\%, 11-20\%, 21-30\%, 31-40\%, 41-50\%). This is presented in Figure
 \ref{A}A. The distribution in baseline recurrence risk appears balanced between the patients who received RT and those who did not; the exception is patients with a baseline local recurrence risk of 31-40\% and 41-50\% (risk buckets 4 and 5, respectively).  To remedy these imbalances, we merged the two strata and oversampled the underrepresented patients such that the sample size of this merged stratum was similar to that of the other risk strata. As shown in Figure \ref{A}B, the imbalanced distribution of baseline recurrence risk between the treated and untreated patients was remedied. Given the lack of confounding, we did not engage the second part of our methodology by introducing the weight $\rho$ and 
 proceeded directly with training a survival RF model in the patients who received RT and a separate survival RF model in the patients who did not.

\begin{figure}[hbtp]
\centering
\caption{Optimal Policy Tree for local recurrence-free survival in patients with extremity soft tissue sarcoma.}
\label{fig5}
\includegraphics[width=0.9\linewidth]{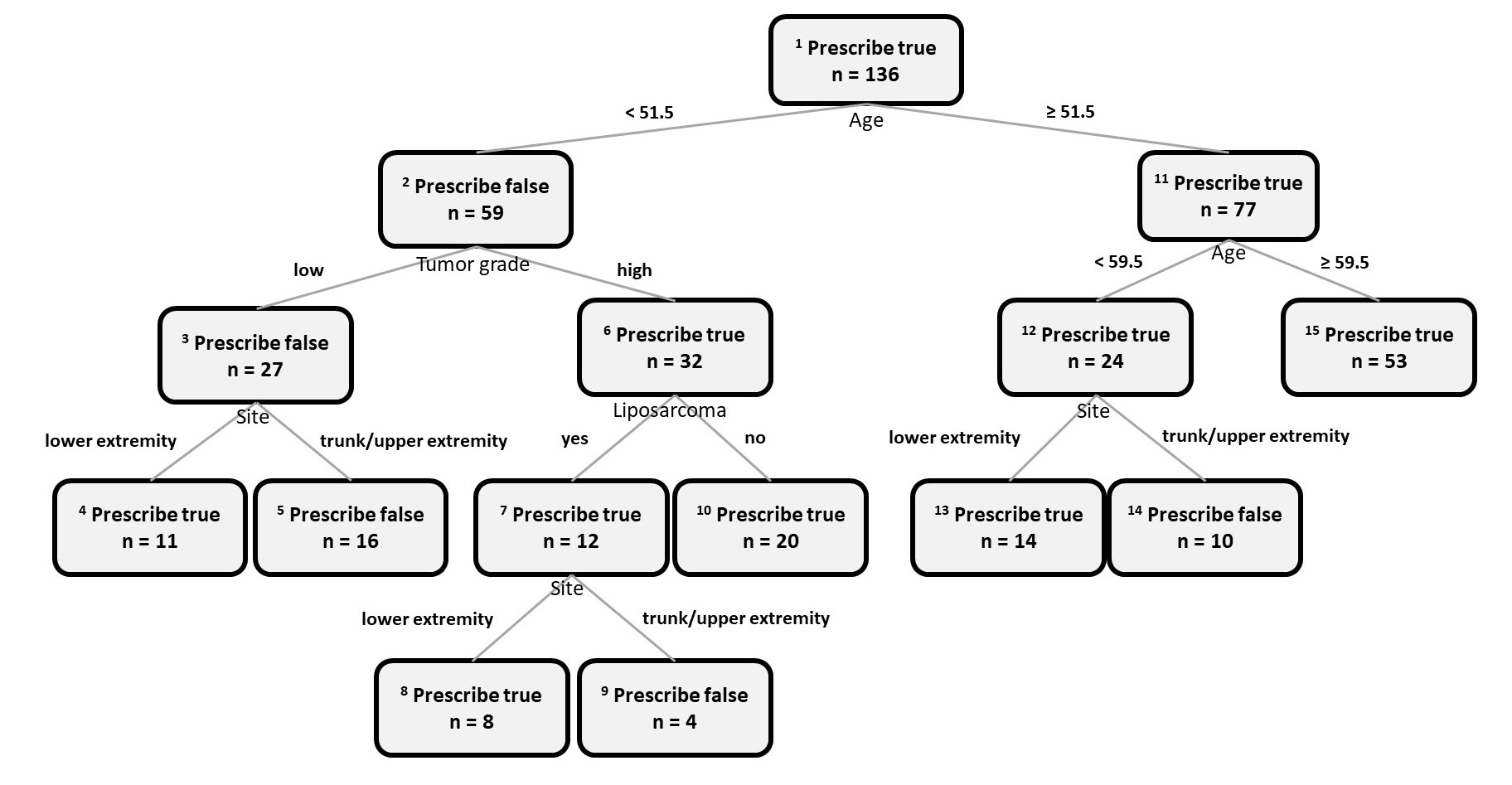}

\end{figure}

\noindent\textit{OPT}\\
\noindent Figure \ref{fig5} illustrates the OPT that was trained using the predictions of the RF survival models. Its hyperparameters were a minbucket of 2 (as the cohort was small) and a max depth of 5. The OPT recommended RT for three patient subgroups and no RT for the other three subgroups. The subgroups for which treatment was recommended included patients older than 60 years, those with lower extremity sarcomas regardless of other characteristics, and those with high grade tumors (with exception to those with liposarcomas). In contrast, adjuvant RT was not recommended for patients with truncal or upper extremity sarcomas (with exception to those with high grade non-liposarcomas) and patients older than 60 years.

%\newpage
\noindent\textit{Validation cohort}\\
\noindent The validation cohort included 936 patients who underwent surgery at MSK between 1982 and 2018 and did not receive RT. The characteristics of these patients are presented in Table \ref{tab3}. At a median follow-up of 76 months (IQR: 41–116), 92 of 936 patients (9.8\%) developed a local recurrence. The 5-year RFS rate was 91\% (95\%CI: 89-93\%), which is better than that of untreated patients in the training cohort. This is expected since no patients who received RT are included in this cohort, as the receipt of RT modifies a patient’s baseline recurrence risk. Importantly, we can use sensitivity and specificity as validation metrics since the prevalence of the outcome (i.e., recurrence) does not influence these metrics.

The sensitivity and specificity of the OPT cut offs was 90\% (95\%CI: 82-95\%) and 14\% (95\%CI: 12-17\%), respectively. The high sensitivity means that the OPT will suggest adjuvant RT to almost all patients who will recur. The low specificity is not an issue since the current treatment recommendation is to offer RT to all these patients. Thus, the OPT can spare around 15\% of patients who would not have benefited from adjuvant RT from an unnecessary treatment. Notably, the authors of the two RCTs on this topic acknowledged that there may exist a subset of patients who do not benefit from adjuvant RT despite their studies findings. In fact, the investigators of the National Cancer Institute (NCI) RCT noted that “although postoperative radiation therapy is highly effective in preventing [local recurrences], selected patients with extremity soft tissue sarcoma may not require adjuvant XRT after limb-sparing surgery” \cite{yang1998randomized}. The authors of the other RCT (the MSK RCT) even tried to identify subsets of patients who may not benefit from RT by using the one-variable-at-a-time analysis \cite{pisters1996long}. They found that RT benefited patients with high grade but not low grade sarcomas. Interestingly, our OPT similarly recommends RT only for patients with high grade tumors in one of its intermediate nodes. However, the OPT further refines these recommendations by suggesting that patients with high grade sarcomas of the upper extremity or trunk do not benefit from radiotherapy, while those with low grade tumors of the lower extremity do. Thus, anatomic site may be the factor that determines which low grade tumors will benefit from RT.  Notably, 53\% of the low grade tumors were found in the lower extremity in the MSK RCT that reported that RT did not benefit patients with low grade sarcomas as compared to 72\% found in the NCI RCT \cite{yang1998randomized}. Interestingly, the investigators of the latter also performed a one-variable-at-a-time sub analysis using tumor grade but found that RT was beneficial in both high and low grade tumors, contradicting the findings of the first RCT. It is possible that the differing frequencies of lower extremity low grade tumors between the two RCTs is the reason behind the disparate findings. This serves as a notable illustration of the human mind's ability to leverage clinical experience for pattern recognition based on a single characteristic, such as distinguishing between low and high-grade tumors, with reasonable success. However, it also highlights the inherent limitations, as the human mind struggles to simultaneously consider multiple variables. Consequently, the pattern recognition based on a single characteristic may lead to substantial oversimplification of complex phenomena.

The investigators of the MSK RCT also performed a univariate analysis to investigate what factors were predictive of local recurrence and found that age greater than 60 years was the only significant predictor \cite{pisters1996long}. This was remarkably consistent with the OPT recommendations, which suggested that RT should be offered to patients older than 59.5 years regardless of other patient characteristics.

Finally, the recommendations of the various OPTs that were trained by tuning the hyperparameters were robust with regard to their sensitivity and specificity, lending credibility to our methodology. Specifically, the sensitivity ranged from 88 to 90\% while the specificity ranged from 14 to 21\%.

\section{Discussion}

In the first part of the study, we applied our novel approach to an observational dataset and matched for observed confounders within the baseline prognostic strata. This allows us to identify a subset of patients that emulates a randomized trial cohort in that the treated and untreated patients within each stratum have in theory similar baseline prognosis. Rosenbaum has previously described a similar prognostic matching approach which reduced heterogeneity and thus reduced sensitivity to unobserved bias \cite{rosenbaum2005heterogeneity}. However, in our example, the counterfactual models that were trained after prognostic matching estimated a higher probability of recurrence under treatment versus no treatment; this cannot be attributed to imatinib since it does not increase the risk of recurrence. Thus, there is residual unobserved confounding that would not be present in a randomized trial. This example emphasizes that even with appropriate matching, unobserved confounding likely exists and “unconfoundness” assumptions may not hold.

We took a second step to account for the residual unobserved confounding; to our knowledge, this step has not been previously suggested. We acknowledge that “capturing” unobserved confounding is an elusive goal since it can be neither observed nor measured. Fortunately, we only need to cancel its effects. First, we used the difference in the probabilities of the outcome (i.e., recurrence) under treatment versus no treatment to quantify the effect of the minimum amount of residual unobserved confounding. We then addressed this by increasing the weight of patients who received the treatment and did not have recurrence, with the aim of equalizing the probabilities of recurrence under treatment versus no treatment.  However, in reality, even a marginally effective treatment will confer slightly better outcomes compared to no treatment at all. Thus, to quantify the effect of the remaining residual unobserved confounding, we serially increased the weight of patients who received the treatment and did not have recurrence and then trained the OPTs using the probabilities of recurrence under treatment and no treatment. The sensitivity and specificity of these OPTs were then compared to identify the weight that best balanced the two metrics. We believe that the optimal weight can address the effect of the remaining residual unobserved confounding by appropriately adjusting the recurrence risk estimates under treatment.

Our example showcases one type of confounding bias that may be present in an observational study; specifically, patients who undergo treatment may have worse outcomes not because the treatment was detrimental but because their baseline risk was higher than that of the untreated patients. However, there is another type of confounding bias that falsely improves outcomes \cite{agoritsas2017adjusted}. For example, patients who receive Treatment A may fare better than those who receive Treatment B due to the presence of favorable but unobserved confounders in the former (e.g., younger age, fewer comorbidities, treatment at high-volume hospitals, etc.); thus, it is the confounders and not the treatment itself that improve patient outcomes. Notably, our approach can address unobserved confounding in this scenario as well. Specifically, we can increase the weight of the patients who received Treatment B and did not have a negative outcome until we find the optimal weight.

To make accurate HTE predictions on an individual level, it is important to address the challenges associated with unobserved confounding. To facilitate clinical implementation, it is equally important that the subsequent OPTs define patient subgroups with heterogeneous treatment effects based on their distinct characteristics. This is highly advantageous as these recommendations can be readily tested for clinical intuitiveness. In contrast, the methodologies that compare treatment effects across different risk strata lack an output in the form of patient subgroups with distinct characteristics. Another weakness of these methodologies relates to the somewhat arbitrary splits that define the different risk strata (e.g., risk deciles) \cite{rekkas2023standardized}. Even when there are varying treatment effects across different strata, the treatment effects within each risk stratum may also vary, making it challenging to implement precision medicine. In contrast, the OPTs split on the level of which variables to include, which cut offs to select, and how the different variables are combined without any user intervention.

Interestingly, hyperparameter tuning of OPTs, which is the only level in which the user is involved, did not result in drastically different OPTs. In fact, regardless of the differences in the structure (e.g., different variables or different variable cut offs), the OPTs are robust in that their sensitivity and specificity converge. This may mean that some patient characteristics used by the OPTs are surrogates of the outcome of interest and not causally related. However, this does not decrease their practical value as the essence of precision medicine is to find the best match between the available treatments and patient subgroups and not the discovery of causal relationships. Importantly, not only did the sensitivity and specificity of the various OPTs converge, but they were in combination higher than those of the current recommendations, which were based on a combination of expert opinion and statistical analysis \cite{miettinen2006gastrointestinal, joensuu2008risk}. Furthermore, the use of sensitivity and specificity is a novel addition as the use of these metrics was previously restricted to validating diagnostic tests only.

Importantly, the value of this framework in observational data can be indirectly validated by comparing the OPT recommendations with those generated by another OPT we previously developed and published \cite{bertsimas2023lancet}. In contrast to the OPT presented in this manuscript, which included all GIST patients regardless of their baseline risk of recurrence, the previous OPT was trained using a cohort limited to patients with intermediate to high risk for recurrence, as determined by the AFIP Miettinen criteria \cite{miettinen2006gastrointestinal}. The reason for this choice is that, at the time of that publication, the methodology presented in this manuscript had not yet been developed. Therefore, to mitigate potential confounding bias, we limited our analysis to high-risk patients who either received or did not receive imatinib, thus ensuring that both groups had comparable high baseline risks of recurrence. Notably, the recommendations of the two OPTs exhibit remarkable consistency. For instance, when the previous OPT recommended adjuvant imatinib for a patient, the new OPT made the same recommendation for 98.7\% of cases. Furthermore, when compared in a validation cohort of high-risk patients, the sensitivity of the new OPT was similar to that of the previous version, with a rate of 95.18\% (95\% CI: 91.73\% to 97.49\%) as opposed to the former OPT's 92.41\% (95\% CI: 88.26\% to 95.44\%). This underscores the validity of the methodology proposed in this manuscript. What is even more significant is that the methodology we introduce in this paper is versatile and applicable in a wide range of clinical contexts, unlike the prior approach, which was inherently limited to a specific subset of patients.

In the second part of this study, we applied the same framework to RCT data. In addition to estimating the ATE in a trial cohort, it is clinically important to also estimate HTE. Currently, HTE is estimated using the conventional approach of one-variable-at-a-time sub-analysis. Specifically, the RCT cohorts are stratified by one variable into two or more subgroups, and the treatment effects within each subgroup are assessed and compared. A fundamental issue with this approach is the inability to identify patient subsets with varying treatment effects when those subsets are defined by a combination of variables \cite{kent2018personalized}. For example, an analysis that stratifies a cohort by mitotic count only will not account for tumor size. This is problematic in a scenario in which a mitotic count of less than 10 and a tumor size of less than 5 cm define a subset of patients who would not benefit from treatment, while a tumor size of greater than 5 cm defines patients who would benefit from treatment. Specifically, if the majority of patients have a tumor size of less than 5 cm, the analysis will erroneously conclude that all patients with a mitotic count of less than 10, even those with a tumor size of greater than 5cm, do not benefit from the treatment. Thus, dividing patients based on this single characteristic (e.g., mitotic count of $\leq10$ vs $>10$) may be grossly misleading as it ignores the heterogeneity present across other tumor characteristics. Unfortunately, even new ML-based approaches use the same logic of a one-variable-at-a-time analysis to output patient subsets defined by a single characteristic \cite{oikonomou2022individualising}.

Interestingly, when we applied our framework to an RCT cohort and stratified extremity sarcoma patients by baseline prognosis, we found imbalances in the distribution of treated and untreated patients in the higher risk strata. This was likely caused by the relatively small sample size of the study, as smaller RCTs or those without complete blocking may have such imbalances. In such cases, prognostic stratification (Step 1) can be used to diagnose prognostic imbalances, which theoretically should not exist in RCT cohorts. In this example, we attempted to train counterfactual models and OPTs despite the distribution imbalances, but the trees could not identify any patient subgroups with HTE. However, after we restored the distribution balance by oversampling the smaller patient group, the new OPTs were able to define patient subgroups with varying treatment effects based on their distinct characteristics. As described in the results section, the OPT recommendations were clinically intuitive, and we were able to identify a subset of patients with distinct characteristics who may not need radiotherapy. This is significant because RCTs have shown an overall benefit of radiotherapy, leading to the current guideline recommending radiotherapy for all patients in a non-individualized manner. Notably, the OPT recommendations were found to have high sensitivity in an external cohort from MSK.

The OPT recommended that only around 20\% of patients should not receive radiotherapy, which is somewhat predictable as, in our analysis, the mean recurrence-free survival estimates for the treated patients were much higher than those for the untreated patients. This may mean that most patients benefit from the treatment and/or that the treatment effect has a high magnitude. In contrast, in the observational data example, the OPT recommended that 35\% of patients should not receive imatinib. This means that the likelihood of identifying patient subgroups with different treatment effects is the highest when the patient groups have similar average treatment effects, which happens in negative trials. This is important because the results of a negative trial may be used as proof that a given treatment is useless. Thus, we can use OPTs to explore an RCT dataset and discover patient subgroups that may benefit from what was previously deemed a “useless” treatment.

In conclusion, we have introduced a novel approach to address the key deficiencies when utilizing observational data and randomized trials to inform precision medicine. Our approach results in the generation of clinically intuitive treatment recommendations, which were rigorously validated in external cohorts. By evaluating these recommendations based on sensitivity and specificity, we demonstrated their superior performance compared to expert-derived recommendations.

Specifically, our approach countered the effects of unobserved confounding in observational data by correcting the estimated probabilities of the outcome under a treatment through a novel two-step process.  In turn, these probabilities served as the basis for training our OPTs. 
OPTs are decision trees that define patient subgroups with distinct treatment effects based on patient characteristics. This process allowed us to generate clinically meaningful treatment recommendations, which were  successfully validated in an external cohort. We extended this framework to an RCT cohort as the trial's limited sample size led to uneven distributions of treated and untreated patients in the high-risk stratum. Once our framework identified such imbalances and remedied them, the OPT identified a subset of patients with distinct characteristics who may not require adjuvant treatment. These recommendations were successfully validated in an external cohort from MSK. We hope that the proposed framework paves the way for precision medicine.

\section*{Acknowledgements}
The authors would like to thank Dr. Samuel Singer and Dr. Murray F. Brennan from the Memorial Sloan Kettering Cancer Center in New York, USA and Dr. Piotr Rutkowski from the Maria Sklodowska-Curie National Research Institute of Oncology in Warsaw, Poland for sharing their datasets. 

\bibliographystyle{apacite}
\bibliography{ref}

\end{document}

% \title{\centering{\Large{\textit{Design and implementation of a standardized framework to allow the use of both observational and randomized data for precision medicine purposes}}}\setstretch{1.5}}

\title{\Large \textit{Design and implementation of a standardized framework to allow the use of both observational and randomized data for precision medicine purposes}
}

\author{Dimitris Bertsimas$^{1}$, Angelos Koulouras$^{1}$, Georgios Antonios Margonis$^{*1,2}$
}

% \authorrunning{Short form of author list} % if too long for running head

\institute{Dimitris Bertsimas \at
              \email{dbertsim@mit.edu} 
           \and
           Angelos Koulouras \at
              \email{angkoul@mit.edu} 
            \and
           Georgios Antonios Margonis, MD, PhD; (Corresponding Author) \at
              \email{margonig@mskcc.org}
            \and
            1  Sloan School of Management and Operations Research Center, E62-560, Massachusetts Institute of Technology,
                MA, 02139 \\
                \\
            2  Department of Surgery, Memorial Sloan Kettering Cancer Center, New York, NY, 10065
}

\date{Received: NA / Accepted: NA}

\maketitle

\section{Limitations}

There are many options underlying classification model $g_{\boldsymbol{\theta}}$. We usually use RF but any model that is granular enough will do. Decision trees offer less granular probabilities: a decision tree with $l$ leaves will provide only $l$ different probabilities. In that case, each leaf can correspond to a bucket. \\

If the risk scores of the model are heavily skewed towards either zero or one, we can use cost-sensitive loss functions and increase the weight of the minority class in the loss function of the model. \\

Generally, models with high accuracy and AUC metrics will construct more accurate buckets. In this case, AUC may be the more relevant metric, because we are mainly interested in the order of the risks. Note that, the performance of these models is also limited by confounding. \\

The width of the buckets and the number of buckets are heavily dependent on the problem and the data at hand. It is rarely the case that each bucket $k$ will have a large number of patients from each treatment group. \\

If one of the treatment groups has very few patients, then we suggest using all the patients in the matching to reduce the loss in modeling power. For example, if there are 30,000 patients with $t=0$ and 2,500 patients with $t=1$, it is likely that in each bucket there will be fewer patients with $t=1$. In this case, we suggest to select 2,500 patients with $t=0$. \\

The previous approach is also recommended if we want to model the original distribution of the patients. We may want to give larger weight to the buckets that originally have the most patients, which are the patients who we expect to see in practice. \\

In practice, high-risk patients are those that are most commonly treated, while low-risk patients are assigned observation. So, in the high-risk buckets the patients with $t_{i}=0$ are relatively few, while in the low-risk buckets the patients with $t_{i}=1$ are relatively few. \\

The counterfactual models can are also up to the practitioner. We recommend using survival models for survival outcomes, especially when there is censoring in the data. \\

Our approach is based on improving the reward estimation process. So, the OPTs are not required for assigning treatments. We can also use regress and compare or other methods for optimally assigning treatments.